\documentclass[runningheads]{llncs}
\usepackage[utf8]{inputenc}
\usepackage[T1]{fontenc}
\usepackage{cite}
\usepackage{amsmath,amssymb,amsfonts}
\usepackage{algorithmic}
\usepackage{graphicx}
\usepackage{setspace}
\usepackage[margin=8pt,skip=5pt,belowskip=0pt,font={small,stretch=0.9},labelfont=bf]{caption}%
\usepackage{subcaption}
\usepackage{textcomp}
\usepackage{booktabs}
\usepackage{multirow}
\usepackage[usenames,dvipsnames,svgnames,x11names,table]{xcolor}
\usepackage[abbreviations]{foreign}  %
\usepackage[binary-units,per-mode=symbol]{siunitx}
\usepackage{ifthen}
\usepackage{paralist}
\usepackage{fontawesome}
\usepackage{url}
\usepackage{pifont}%
\usepackage[colorlinks=true,linkcolor=blue,urlcolor=blue,citecolor=blue,bookmarks=false]{hyperref}%

\graphicspath{{./}{figures/}}

\newcommand{\tz}{\textsc{TrustZone}\xspace}
\newcommand{\arm}{\textsc{Arm}\xspace}
\newcommand{\ARM}{\arm} %
\newcommand{\optee}{\textsc{Op-Tee}\xspace}
\newcommand{\gp}{\textsc{GlobalPlatform}\xspace}
\newcommand{\posix}{\textsc{Posix}\xspace}

\newcommand{\sys}{\textsc{iperfTZ}\xspace}

\newboolean{showcomments}
\setboolean{showcomments}{false}
\ifthenelse{\boolean{showcomments}}
{ \newcommand{\mynote}[3]{
   \fbox{\bfseries\sffamily\scriptsize#1}
   {\small$\blacktriangleright$\textsf{\emph{\color{#3}{#2}}}$\blacktriangleleft$}}}
{ \newcommand{\mynote}[3]{}}

\newcommand{\vs}[1]{\mynote{Valerio}{#1}{blue}}

\definecolor{darkgreen}{rgb}{0.3,0.5,0.3}
\definecolor{darkblue}{rgb}{0.3,0.3,0.5}
\definecolor{darkred}{rgb}{0.5,0.3,0.3}

\begin{document}

\title{\sys: Understanding Network Bottlenecks for TrustZone-based Trusted Applications\\
[0mm]\normalsize{(Regular Submission, Track D. Security and Privacy)}\thanks{This is a pre-print of an article published in "Stabilization, Safety, and Security of Distributed Systems" (SSS) 2019. The final authenticated version is available online at: \url{https://doi.org/10.1007/978-3-030-34992-9_15}}}

\date{}

\author{Christian~Göttel \and
Pascal~Felber \and
Valerio Schiavoni}

\titlerunning{\sys: Understanding Network Bottlenecks for TrustZone} %
\institute{
  University of Neuch\^atel\\
  Rue Emile-Argand 11, CH-2000 Neuch\^atel, Switzerland\\
\email{first.last@unine.ch}\\
+41~32~718~2722
}

\maketitle
\begin{abstract}
The growing availability of hardware-based trusted execution environments (TEEs) in commodity processors has recently advanced support (\ie design, implementation and deployment frameworks) for network-based secure services.
Examples of such TEEs include \arm~\tz or Intel SGX, largely available in embedded, mobile and server-grade processors.
TEEs shield services from compromised hosts, malicious users or powerful attackers.
TEE-enabled devices are largely being deployed on the edge of the network, paving the way for large-scale deployments of trusted applications.
These applications allow processing and disseminating sensitive data without having to trust cloud providers.
However, uncovering network performance limitations of such trusted applications is difficult and currently lacking, despite the interest and reliance by developers and system deployers.

\sys is an open-source tool to uncover network performance bottlenecks rooted at the design and implementation of trusted applications for \arm~\tz and underlying runtime systems.
Our evaluation based on micro-benchmarks shows current trade-offs for trusted applications, both from a network as well as an energy perspective; an often overlooked yet relevant aspect for edge-based deployments.
\end{abstract}

\vspace{-10pt}
\section{Introduction}
Services are being moved from the cloud to the edge of the network.
This migration is due to several reasons: lack of trust on the cloud provider~\cite{baumann2015shielding}, energy savings~\cite{lyu2018selective,ning2019green} or willing to reclaim control over data and code.
Edge devices are used to accumulate, process and stream data~\cite{makinen2015streaming,varghese2016challenges}.
The nature of such data can be potentially very sensitive: edge devices can be used to process health-based data emitted by body sensors (\eg cardiac data~\cite{segarra2019}), data originated by smart home sensors indicating the presence or absence of humans inside a household, or even financial transactions~\cite{lind2016teechan,shepherd2017establishing}.
In this context, applications using this information must be protected against powerful attackers, potentially even with physical access to the devices.
Additionally, communication channels for inter-edge device applications must also be secured to prevent attacks such as man-in-the-middle attacks.
Edge devices are typically low-energy units with limited processing and storage capacity.
As such, it is unpractical to rely on sophisticated software-based protection mechanisms (\eg homomorphic encryption~\cite{naehrig2011can}), currently due to their high processing requirements and low performance~\cite{gottel2018security}.
Alternatively, new hardware-based protection mechanisms can be easily leveraged by programmers to provide the mentioned protection guarantees.
Specifically, \emph{trusted execution environments} (TEEs) are increasingly made available by hardware vendors in edge-devices~\cite{shepherd2016secure}.
Several \ARM-based devices, such as the popular Raspberry Pi\footnote{\url{https://www.raspberrypi.org}, accessed on 30.07.2019}, embed native support for \tz~\cite{ngabonziza2016trustzone,arm:tz}, \ARM's specific design for TEEs.
\tz can be leveraged to deploy \emph{trusted applications} (TAs) with additional security guarantees. 
There exist several programming frameworks and runtime systems to
develop TAs for \tz with varying capabilities and different degrees of
stability and support (\eg SierraTEE\footnote{\url{https://www.sierraware.com/open-source-ARM-TrustZone.html}, accessed on 30.07.2019}, \optee\footnote{\url{https://www.op-tee.org}, accessed on 30.07.2019}, and~\cite{McGillion2015OpenTEEA}).
While few studies look at the interaction between TEEs and the corresponding untrusted execution environments~\cite{jang2015secret,amacher2019}, little is known on the network performance bottlenecks experienced by TAs on \arm processors. %
We fill this gap by contributing \sys, a tool to measure accurately the network performance (\eg latency, throughput) of TAs for \tz.
\sys consists of three components, namely \textit{(1)} a client application, \textit{(2)} a TA, and \textit{(3)} a server.
Our tool can be used to guide the calibration of TAs for demanding workloads, for instance understanding the exchanges with untrusted applications or for secure inter-TEE applications~\cite{shepherd2017establishing}.
In addition, \sys can be used to study the impact of network and memory performance on the energy consumption of running TAs. 
By adjusting \sys's parameters, users evaluate the network throughput of their TAs and can quickly uncover potential bottlenecks early in the development cycle.
For instance, internal buffer sizes affect the achievable network throughput rates by a factor of \SI{1.8}{\times}, almost halving throughput rates.

The rest of the paper is organized as follows.
\autoref{sec:usecase} motivates the need for tools analyzing TAs. 
We provide an in-depth background on \tz in \autoref{sec:background}, as well as covering details on the \tz runtime system \optee.
In \autoref{sec:archicture} we present the architecture of \sys and some implementation details in \autoref{sec:implementation}. %
We report our evaluation results in \autoref{sec:evaluation}.
We cover related work in \autoref{sec:rw} before concluding in \autoref{sec:conclusion}.

\section{Motivating Scenario}
\label{sec:usecase}
We consider scenarios with simple yet practical services deployed as TAs.
For instance, in~\cite{dais19} authors deploy key-value stores inside a \tz runtime system.
Benchmarks show a $12\times$-$17\times$ slowdown when compared to plain (yet unsecure) deployments, due to shared memory mechanisms between the trusted and untrusted environments.
As further detailed in \autoref{sec:archicture}, networking in \optee is supported by similar shared memory mechanisms. 
Yet, we observe the lack of tools to clearly highlight the root causes of such bottlenecks.
Further, in the \tz ecosystem, there is a lack of proper tools to evaluate network bottlenecks contrary to untrusted environments (\eg, \texttt{iperf3}\footnote{\url{https://software.es.net/iperf/}, accessed on 30.07.2019}, \texttt{netperf}\footnote{\label{fn:netperf}\url{https://hewlettpackard.github.io/netperf/}, accessed on 30.07.2019}, \texttt{nuttcp}\footnote{\label{fn:nuttcp}\url{https://www.nuttcp.net/Welcome\%20Page.html}, accessed on 30.07.2019}).
The overhead originating from the shared memory mechanism can be identified by comparing the measured network throughput inside and outside the TEE. 
Measuring such overheads is of particular relevance in embedded, mobile and IoT environments. 
In those scenarios, devices are often battery powered, limited both in time and capacity. %
Hence, network performance tools should further highlight energy costs, pointing users to specific bottlenecks.
\begin{figure*}[t]
  \centering 
  \begin{subfigure}[t]{0.495\textwidth}
    \centering
    \includegraphics[width=\linewidth]{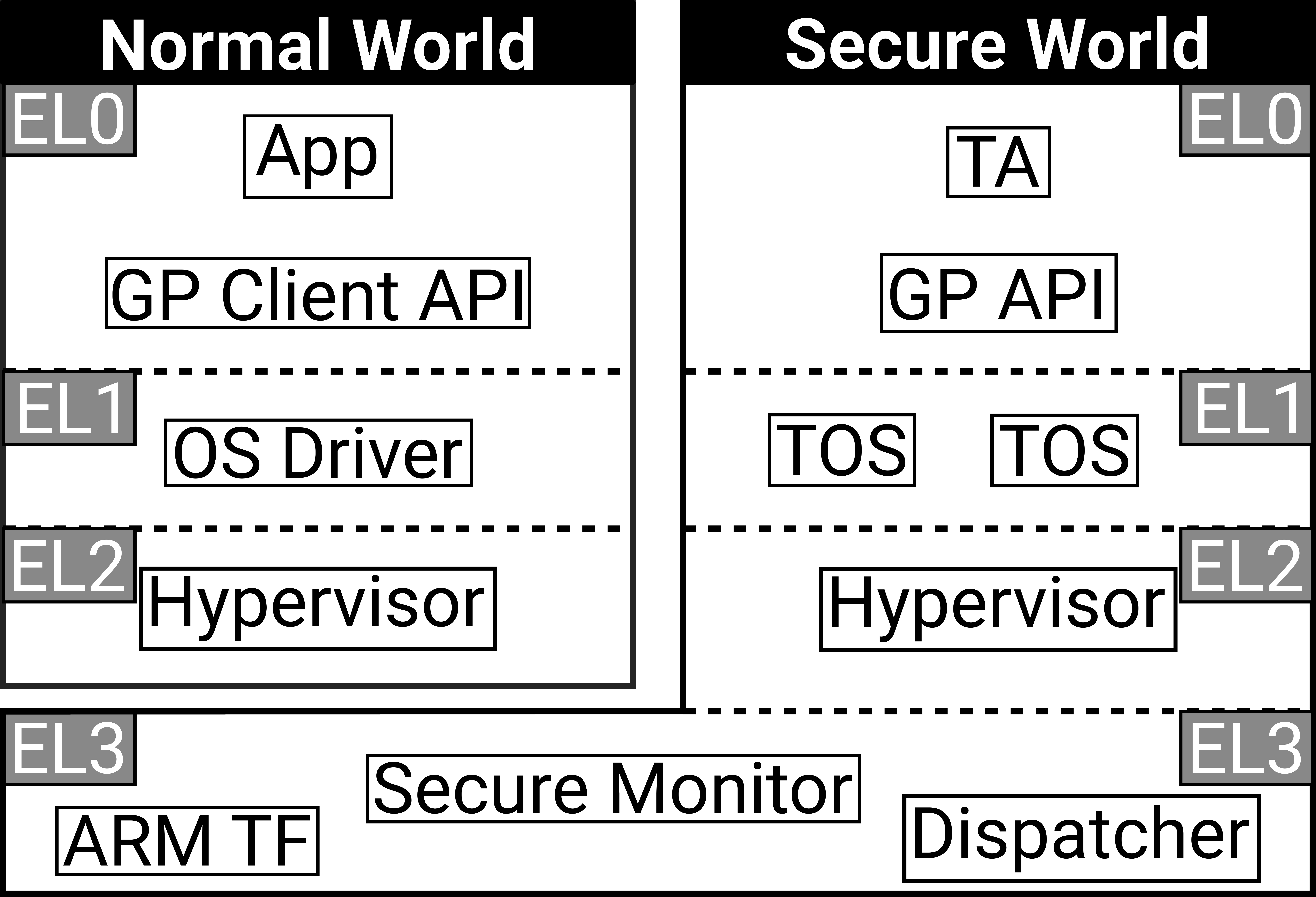}
    \caption{{\arm}v8.4-A architecture~\cite{arm:sel2}}
    \label{fig:blocktz}
  \end{subfigure}
  \quad%
  \begin{subfigure}[t]{0.465\textwidth}
    \centering
    \includegraphics[width=\linewidth]{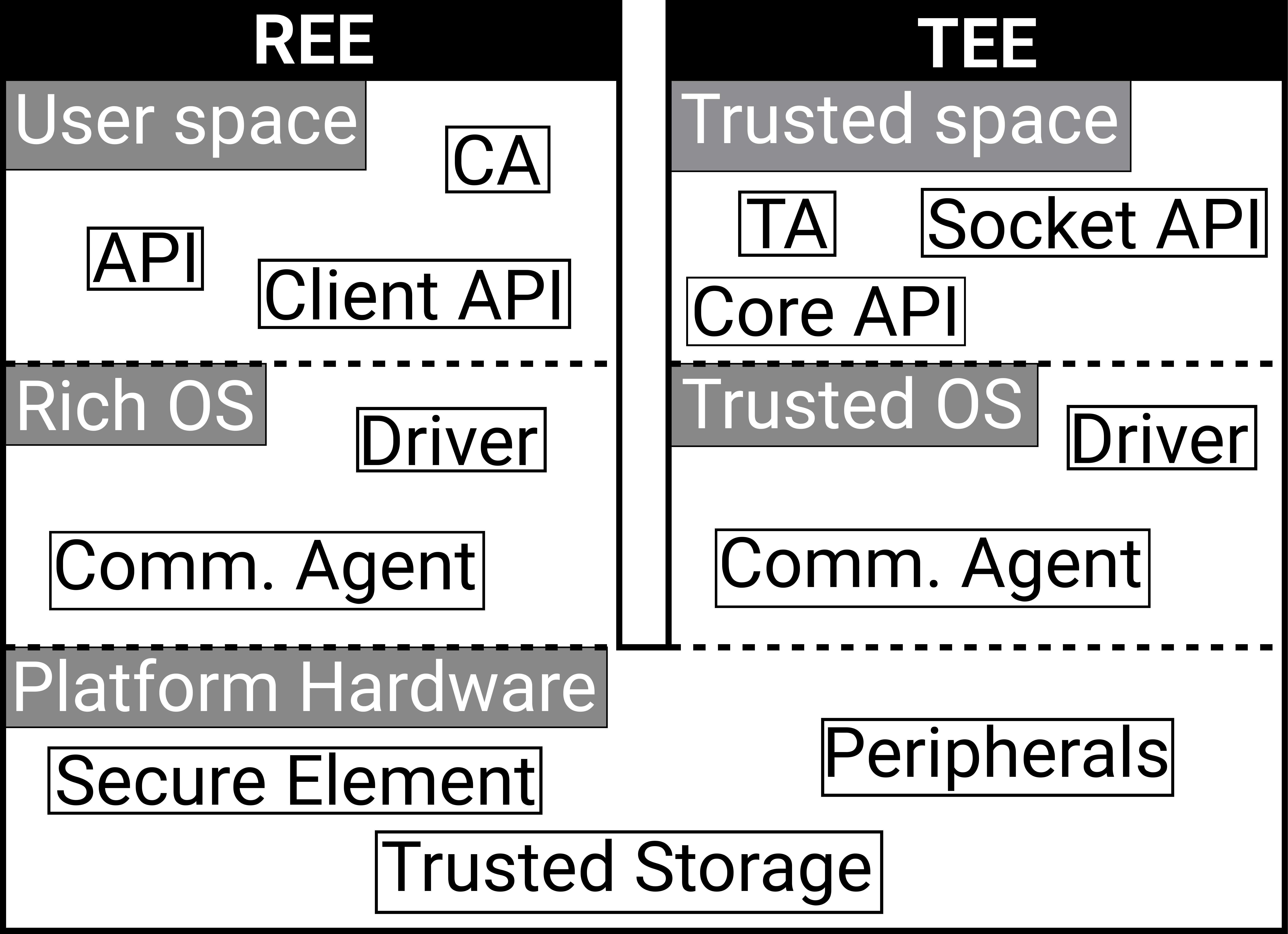}
    \caption{\gp architecture~\cite{gp:sysarch}}%
    \label{fig:blockgp}
  \end{subfigure}%
  \caption{Block diagrams highlighting relevant software components.}%
  \label{fig:blockdia}
\end{figure*}

\section{Background}
\label{sec:background}
This section provides a background on \arm \mbox{\tz} (\autoref{subsec:tz}), the \gp specifications (\autoref{subsec:gpspec}) and \optee, the \tz runtime system used for \sys (\autoref{subsec:optee}).
This background helps understanding technical challenges in our context and how \sys addresses them.

\subsection{ARM TrustZone in a Nutshell}
\label{subsec:tz}
\tz is a security architecture designed for \arm processors and was introduced in 2003~\cite{arm:sel2}.
It partitions hardware and software resources into two worlds, \ie \emph{secure} and \emph{normal} world, as shown in \autoref{fig:blocktz}.
A dedicated \texttt{NS} bit~\cite{arm:tz} drives this world separation and allows to execute secure (\texttt{NS} bit set low) or non-secure (\texttt{NS} bit set high) transactions on the system bus. %
In general, non-secure transactions cannot access system resource secured by a low \texttt{NS} bit.
The \tz architecture spans beyond the system bus, including peripherals (\eg GPUs~\cite{volos2018graviton} and I/O). %
Every \tz-enabled processor is logically split into a secure and a non-secure (virtual) core, executing in a time-shared manner.
Hence, accessible system resources are determined by the executing core: secure cores can access all system resources, while non-secure cores can only access non-secure ones.
\arm processors embed one \emph{memory management unit} (MMU) per virtual core in charge of mapping virtual addresses to physical addresses.
The \emph{translation lookaside buffer} (TLB) in the MMU is used to maintain the mapping translations from virtual to physical memory addresses.
Tagging TLB entries with the identity of the world %
allows secure and non-secure address translation entries to co-exist.
With tags the TLB no longer has to be flushed making fast world switches possible.

The implementation of \tz is organized into four \emph{exception levels} (EL) with increasing privileges~\cite{arm:a53} (\autoref{fig:blocktz}).
EL0, the lowest one, executes unprivileged software. 
EL1 executes operating systems, while EL2 provides support for virtualization. %
Finally, \arm Trusted Firmware is running at EL3 dispatching boot stages at boot time and monitoring secure states.
Switches between the two worlds are supervised by a secure monitor~\cite{arm:v8}. %
It is invoked in two ways: \emph{(1)} by executing a \emph{secure monitor call} (SMC), or \emph{(2)} by a subset of \emph{hardware exception mechanisms}~\cite{arm:tz}.
When invoked, the secure monitor saves the state of the currently executing world, before restoring the state of the world being switched to.
After dealing with the worlds' state, the secure monitor returns from exception to the restored world.

\subsection{The GlobalPlatform Standard}
\label{subsec:gpspec}
\gp\footnote{\url{https://globalplatform.org}, accessed on 30.07.2019} publishes specifications for several TEEs (\eg \optee and~\cite{McGillion2015OpenTEEA}).
We provide more details on \optee in \autoref{subsec:optee} (an implementation of such specifications), while briefly explaining the terminology in the remainder to understand~\autoref{fig:blockgp}.
An \emph{execution environment} (EE) provides all components to execute applications, including hardware and software components. %
A \emph{rich execution environment} (REE) runs a rich OS, generally designed for performance.
However, it lacks access to any secure component.
In contrast, TEEs are designed for security, but programmers have to rely on a reduced set of features.
A trusted OS manages the TEE under constrained memory and storage bounds.
TEE and REE run alongside each other.
In recent \arm releases (since v8.4), multiple TEEs can execute in parallel~\cite{arm:sel2}, each with their own trusted OS.
TAs rely on system calls implemented by the trusted OS, typically implemented as specific APIs~\cite{gp:internal}.
\emph{Client applications} (CA) running in the rich OS can communicate with TAs using the \emph{TEE Client API}.
Similarly, TAs can access resources such as \emph{secure elements} (\ie tamper-resistant devices), \emph{trusted storage}, and \emph{peripherals}, or send messages outside the TEE.
\emph{Communication agents} in the TEE and REE mediate exchanges between TAs and CAs.
Finally, the \emph{TEE Socket API} can be used by TAs to setup network connections with remote CAs and TAs.
\vs{still too many acronyms here, but should be a bit better now}

\subsection{\optee: Open Portable Trusted Execution Environment}
\label{subsec:optee}
\optee is an open-source implementation of several \mbox{\gp}  specifications~\cite{gp:sysarch,gp:client,gp:internal,gp:socket} with native support for \tz.
The \optee OS manages the TEE resources, while any Linux-based distribution can be used as rich OS alongside it.
\optee supports two types of TAs: \emph{(1)} regular TAs~\cite{gp:sysarch} running at EL0, and \emph{(2)} \emph{pseudo TAs} (PTAs), statically linked against the \optee OS kernel.
PTAs run at EL1 as secure privileged-level services inside \optee OS's kernel.
Finally, \optee provides a set of client libraries to interact with TAs and to access secure system resources from within the TEE.
\section{Networking for Trusted Applications}
\label{sec:archicture}

The application in the REE acts as a proxy interface to the TA forwarding arguments.
First, the application creates a context (\texttt{TEEC\_InitializeContext}). 
Then, it allocates dynamic shared-memory areas (\texttt{TEEC\_AllocateSharedMemory}), basically used as buffer between the secure and normal worlds, as well as forward function arguments, piggybacked upon the session creation (\texttt{TEEC\_OpenSession}).
Functions in the TA can be used through \texttt{TEEC\_InvokeCommand} calls.
Once the session is closed  (\texttt{TEEC\_CloseSession}), the shared-memory areas can be released from the context and freed before finalizing the context.

For networked TAs, \ie generating or receiving network traffic
respectively from and to TAs, runtime systems must provide support for sockets and corresponding APIs. %
To do so, either \emph{(1)} the TEE borrows the network stack from the REE, or \emph{(2)} the TEE relies on \emph{trusted device drivers}.
The former solution implies leveraging \emph{remote procedure calls} (RPC) to a \texttt{tee-supplicant} (an agent which responds to requests from the TEE), and achieves a much smaller \emph{trusted computing base}.
The latter allows  for direct access to the network device drivers for much lower network latencies.
Furthermore, it simplifies confidential data handling as the data does not have to leave the TEE.
The former requires developers to provide data confidentiality before network packets leave the TEE, for instance by relying on encryption.

\sys leverages \texttt{libutee}\footnote{\url{https://optee.readthedocs.io/architecture/libraries.html\#libutee}, accessed on 30.07.2019} and its socket API, supporting streams or datagrams.
The socket interface exposes common functions: \texttt{open}, \texttt{send}, \texttt{recv}, \texttt{close}, \texttt{ioctl} and \texttt{error}.
The \gp specification allows TEE implementations to extend protocol-specific functionalities via command codes and \texttt{ioctl} functions. 
For example, it is possible to adjust the receiving and sending socket buffer sizes with TCP socket or changing the address and port with UDP sockets.

The \texttt{libutee} library manages the lifecycle of sockets via a TA session to the socket's PTA.
The socket PTA handles the RPC to the \texttt{tee-supplicant}, in particular allocating the RPC parameters and assigning their values.
Afterwards, a SMC instruction is executed to switch back to the normal world.
The \texttt{tee-supplicant} constantly checks for new service requests from the TEE.
Once a new request arrives, its arguments are read by the \texttt{tee-supplicant} and the specified command is being executed.
Finally, when the data is received by the \texttt{tee-supplicant}, it is relayed over \posix sockets to the rich OS.
In essence, when data is sent or received over a socket, it traverses
all exception levels, both secure (from EL0 up to EL3) and non-secure (from EL2 to EL0 and back up).

\begin{figure*}[t]
  \begin{minipage}{0.48\textwidth}
    \includegraphics[width=\linewidth]{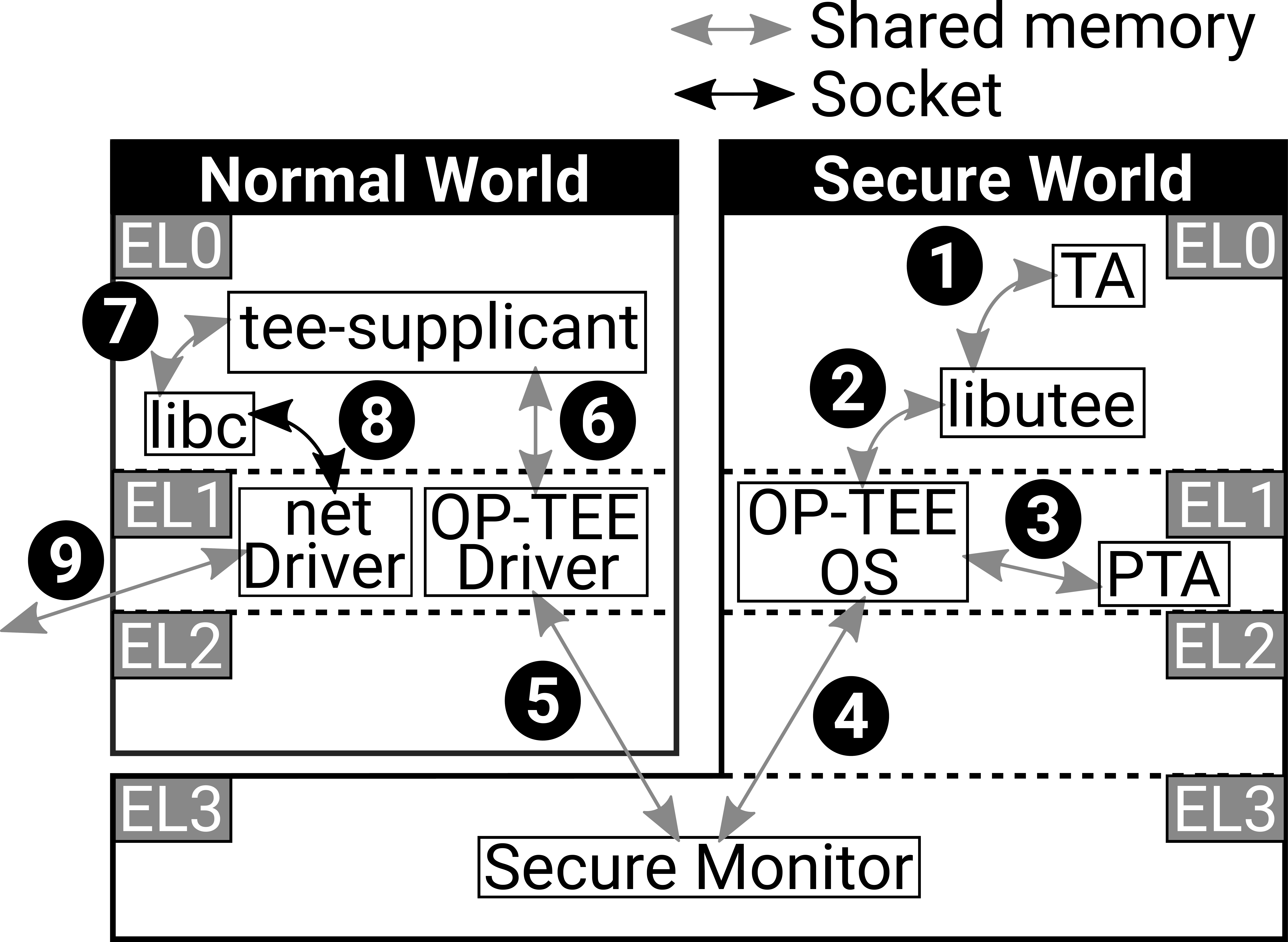}
    \caption{Execution flow inside \optee.}%
    \label{fig:optee}
  \end{minipage}%
  \quad%
  \begin{minipage}{0.48\textwidth}
    \centering
    \includegraphics[width=0.68\linewidth]{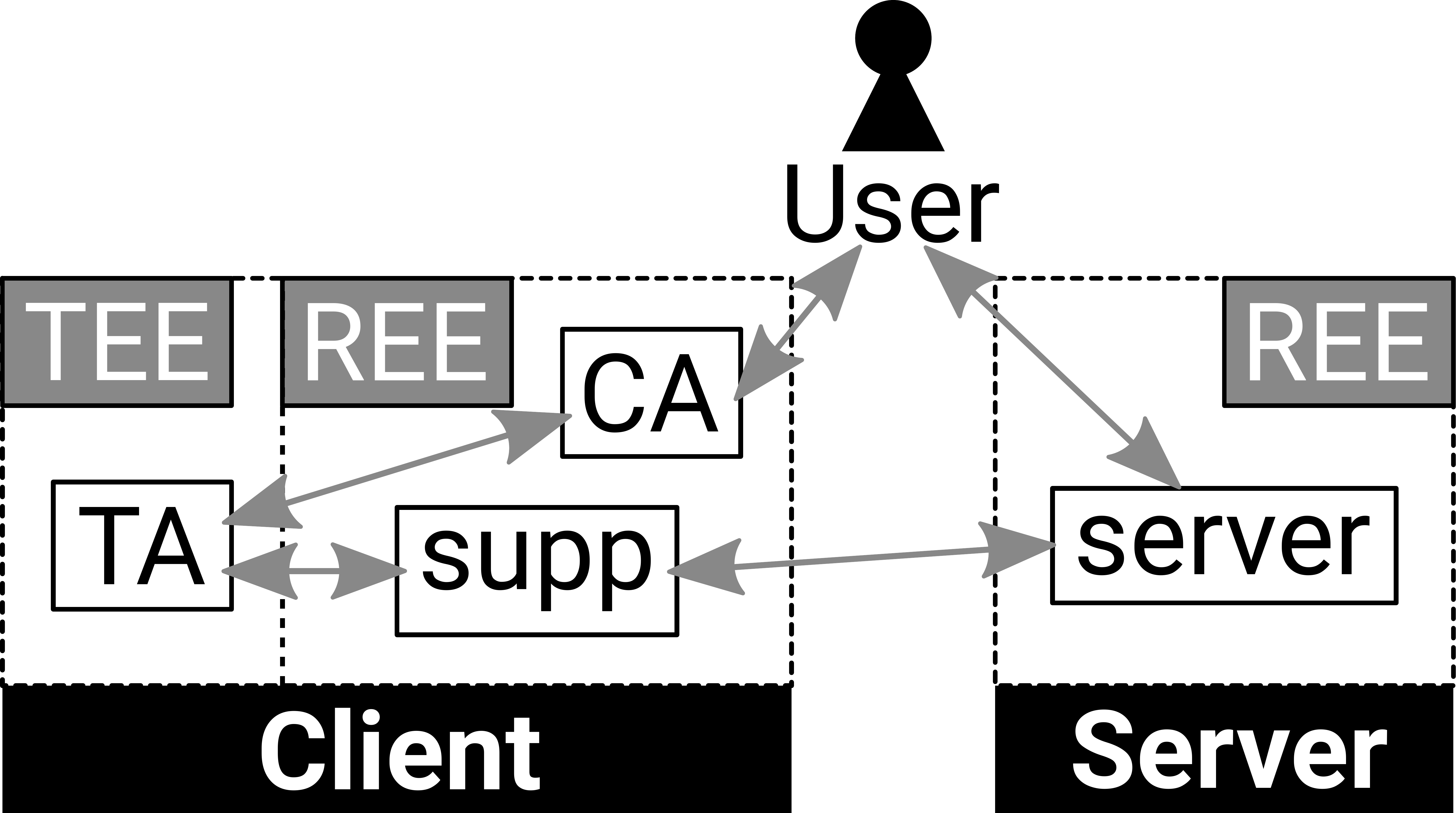}
    \caption{Interaction of \sys's components in the client-server model.}%
    \label{fig:sys}
  \end{minipage}
\end{figure*}

\autoref{fig:optee} summarizes the previous paragraphs and shows the interaction between the secure and normal worlds in \optee.
The secure world hosts the TA, which interacts directly with \texttt{libutee} (\autoref{fig:optee}-\ding{202}).
When using \gp's Socket API, \texttt{libutee} does a system call (\autoref{fig:optee}-\ding{203}) to \optee.
\optee then delegates the request to the socket PTA (\autoref{fig:optee}-\ding{204}).
The secure monitor is invoked through a SMC (\autoref{fig:optee}-\ding{205}), which maps the data from the TEE to the REE's address space.
From there execution switches into the normal world and the \optee driver (\autoref{fig:optee}-\ding{206}) resumes operation.
Requests are then handled by the \texttt{tee-supplicant} (\autoref{fig:optee}-\ding{207}) over \texttt{ioctl} system calls. 
The agent executes system calls using \texttt{libc} (\autoref{fig:optee}-\ding{208}) to directly relate the underlying network driver (\autoref{fig:optee}-\ding{209}) over the \posix interface. 
Once data reaches the network driver, it can be sent over the wire (\autoref{fig:optee}-\ding{210}).

\subsection{Threat Model}
\label{subsec:threatmodel}

For our threat model we consider a malicious user that has physical access or is capable to obtain remote access on the devices used to deploy \sys as depicted in~\autoref{fig:sys}.
By gaining access to the devices or the network the devices are connected to, the malicious user has either the intention to compromise these devices or to exploit \sys for denial-of-service attacks.
We assume that the REE, which includes the rich OS and the user space, cannot be trusted. However, we assume that the devices and the TEE, which includes bootloader, \optee, and secure monitor, can be trusted.
As also stated in~\cite{arm:tz}, side-channel attacks are out of scope of our threat model.

Assuming our \tz-enabled device is equipped with an \emph{embedded MultiMediaCard} (eMMC), then TAs can be securely stored on the eMMC and the malicious user cannot tamper with a TA's binary.
Otherwise, the malicious user who has gained control over the REE, has access to the TAs and can manipulate a TA's binary and compromise \sys.
Manipulation of the CA's parameters by the malicious user to trigger a buffer overflow can be excluded.
In \sys, buffer allocation and initialization use the same variable as size indicator.
Hence, the TA will return an out of memory error code, if it tries to allocate more memory than it is allowed to use.
During a network bandwidth measurement, the malicious user can run a (distributed) denial-of-service attack to reduce the network bandwidth, such that a lower network throughput is measured and reported by \sys.
Although irrelevant to \sys, the malicious user could run a man-in-the-middle attack, either directly within the REE or on the network, and intercept the traffic exchanged between the two devices.
At the time of writing, \optee does not provide support for the TLS protocol which renders secure connections unusable.
\section{Implementation}
\label{sec:implementation}
We describe the implementation challenges of the three components included in \sys,\footnote{The source code of \sys will be made available on GitHub.} namely \textit{(1)} a CA acting as proxy for \sys's \textit{(2)} TA, and \textit{(3)} the server component which the TA is interfacing.
All components are implemented in the C language, and consists of 927 lines of code: 243 for the client, 314 for \sys's TA, and 430 for the server.\footnote{Numbers for individual components include local header lines of code.}

\subsection{\sys: Client Application}
When the CA starts, the TEE context is initialized using the file descriptor fetched from the \optee driver.
Two distinct dynamic shared-memory areas are allocated at this time, to \emph{(1)} exchange arguments passed over the \emph{command line interface} with the TA (see \autoref{subsec:ta}) and \emph{(2)} to retrieve metrics gathered by the TA during the network measurement. %
Several arguments (\eg, IP of the target server node, dummy data size, socket buffer size) are written in the shared memory area.
The dummy data size is used by the TA to read/write data to the interface socket.
Both shared memory areas get registered with the operation data structure before calling the \texttt{TEEC\_InvokeCommand} function.
The executing thread in the CA is blocked until the TA completes. 
The execution inside the TEE is resumed at the TA's main entry point upon world switch.
Once the TA completes, an \texttt{SMC} instruction drives the CPU core to switch back into the normal world, where execution is resumed.
The metrics gathered from the TA are available to the user as persistent files.

\subsection{\sys: Trusted Application}
\label{subsec:ta}
The \sys TA is the primary executing unit. 
It takes the role of the client in the client-server model. %
The TA allocates a buffer for the dummy data on the heap, filled with random data generated by \optee's Cryptographic Operations API~\cite{gp:internal}.
With the information from the arguments, the TA finally sets up a TCP interface socket and opens a client connection before assigning the socket buffer sizes.
Our implementation relies on the Time API~\cite{gp:internal} to measure the elapsed time during the network throughput measurement inside the TEE.
\optee computes the time value from the physical count register and the frequency register.
The count register is a single instance register shared between normal and secure world EL1.
The network throughput measurement is then started while either maintaining a constant bit rate, transmitting a specific number of bytes or running for $10$ seconds.
During the measurement, the TA gathers metrics on the number of transmit calls, \ie \texttt{recv} and \texttt{send}, bytes sent, time spent in the transmit calls and the total runtime.
Upon completion, results are written to the shared memory area and the execution switches back to the normal world.

\subsection{\sys: Server}
\label{subsec:server}
The server component is deployed and executed inside the normal world.
This is used to wait for incoming TCP connections (or inbound UDP datagrams) from \sys's TA.
While executing, it gathers similar network metrics as the other components.
Additionally, this component collects TCP specific metrics, such as the smoothed \emph{round trip time} or the \emph{maximum segment size}.
This TCP specific data is not accessible for TAs and can only be retrieved on the server side using a \texttt{getsockopt} system call.

\section{Evaluation}
\label{sec:evaluation}
In this section we will demonstrate how \sys can measure the network throughput.
We further draw conclusions regarding hardware and software implementation designs.
We report that it is particularly challenging to assess network throughput, given the remarkable diversity one can find on embedded and mobile \arm systems.

\begin{table}[t]
  \footnotesize
  \centering
  \caption{Comparison of evaluation platforms.}%
  \label{tab:platform}
  \setlength{\aboverulesep}{0pt}
  \setlength{\belowrulesep}{0pt}
  \rowcolors{1}{gray!10}{gray!0}
  \begin{tabular}{>{\kern-\tabcolsep}lll<{\kern-\tabcolsep}}
    \toprule\rowcolor{gray!25}
    \multicolumn{1}{c}{\textbf{Device}} & \multicolumn{1}{c}{\textbf{QEMU}} & \multicolumn{1}{c}{\textbf{Raspberry}} \\
    \midrule%
    CPU Model & Intel Xeon E3-1270 v6 & Broadcom BCM2837 \\
    \rowcolor{gray!10}
    CPU Frequency & \SI{3.8}{\GHz} & \SI{1.2}{GHz} \\
    \rowcolor{gray!0}
    Memory Size & \SI{63}{\gibi\byte} DDR4 &
    \SI{944}{\mebi\byte} LPDDR2 \\
    \rowcolor{gray!10}
    Memory data rate & \SI{2400}{\mega T\per\second} & \SI{800}{\mega T\per\second} \\
    \rowcolor{gray!0}
     & Samsung & Transcend micro SDHC \\
    \rowcolor{gray!0}
    \multirow{-2}{*}{Disk Model} & MZ7KM480HMHQ0D3 & UHI-I Premium \\
    \rowcolor{gray!10}
    Disk Size & \SI{480}{\giga\byte} & \SI{16}{\giga\byte} \\
    \rowcolor{gray!0}
    Disk Read Speed & \SI[per-mode=symbol]{528.33}{\mega\byte\per\second} &
    \SI[per-mode=symbol]{90}{\mega\byte\per\second} \\
    Network Bandwidth & \SI[per-mode=symbol]{1}{\giga\bit\per\second} &
    \SI[per-mode=symbol]{100}{\mega\bit\per\second} \\
    \bottomrule
  \end{tabular}
\end{table}

\textbf{Evaluation settings.}
We deploy \sys on the Raspberry Pi platform.
Due to the limited network bandwidth of Raspberry Pi devices supported by \optee, we also include results under emulation using QEMU.\footnote{\url{https://www.qemu.org}, accessed on 30.07.2019}
With QEMU we can run the same evaluation as on the Raspberry Pi and we also profit from a higher network bandwidth.
\autoref{tab:platform} compares in detail the two setups.
For both setups we use the same machine as server, on which we collect power consumptions and run the \sys server component.

\textbf{Server.} The server is connected to a Gigabit switched network, with access to power meter measurements.
The nodes being measured are at a single-hop from the server.
During the micro-benchmarks server components will be deployed on the server with fixed dummy buffer and socket buffer sizes of \SI{128}{\kibi\byte}.
This allows creating an accurate time series of the recorded throughput, latency and power metrics by concentrating the data acquisition on a single node.

\textbf{QEMU.} We deploy \optee with QEMU v3.1.0-rc3 running on a Dell PowerEdge R330 server.
The \optee project has built-in support for QEMU and uses it in system emulation mode.
In system emulation mode QEMU emulates an entire machine, dynamically translating different hardware instruction sets when running a virtual machine with a different architecture.
In order to provide full network capability, we replace the default SLiRP network\footnote{\url{https://wiki.qemu.org/Documentation/Networking\#User_Networking_.28SLIRP.29}, accessed on 30.07.2019} deployed with \optee by a bridged network with a tap device.

\textbf{Raspberry Pi.} %
\optee only supports the Raspberry Pi 3B.
We deploy \optee on a Raspberry Pi 3B v1.2 equipped with a Broadcom BCM2837 SoC.
The SoC implements an ARM Cortex-A53 with ARMv8-A architecture. %
The BCM2837 chip lacks support for cryptographic acceleration instructions and  %
is not equipped with \emph{\tz Protection   Controller} (TZPC), \emph{\tz Address Space Controller} (TZASC), \emph{Generic Interrupt Controller} (GIC) or any other proprietary security control interfaces on the bus~\cite{sequitur}.
The Raspberry Pi 3B lacks an on-chip memory or eMMC to provide a securable memory.
We take these limitations into account in our evaluation, and leave further considerations once a more mature support for the Raspberry Pi platform is released.

\begin{figure*}[ht]
  \centering
  \begin{subfigure}[t]{0.50\textwidth}
    \includegraphics[trim={5px 30px 35px 205px},clip,width=\linewidth]{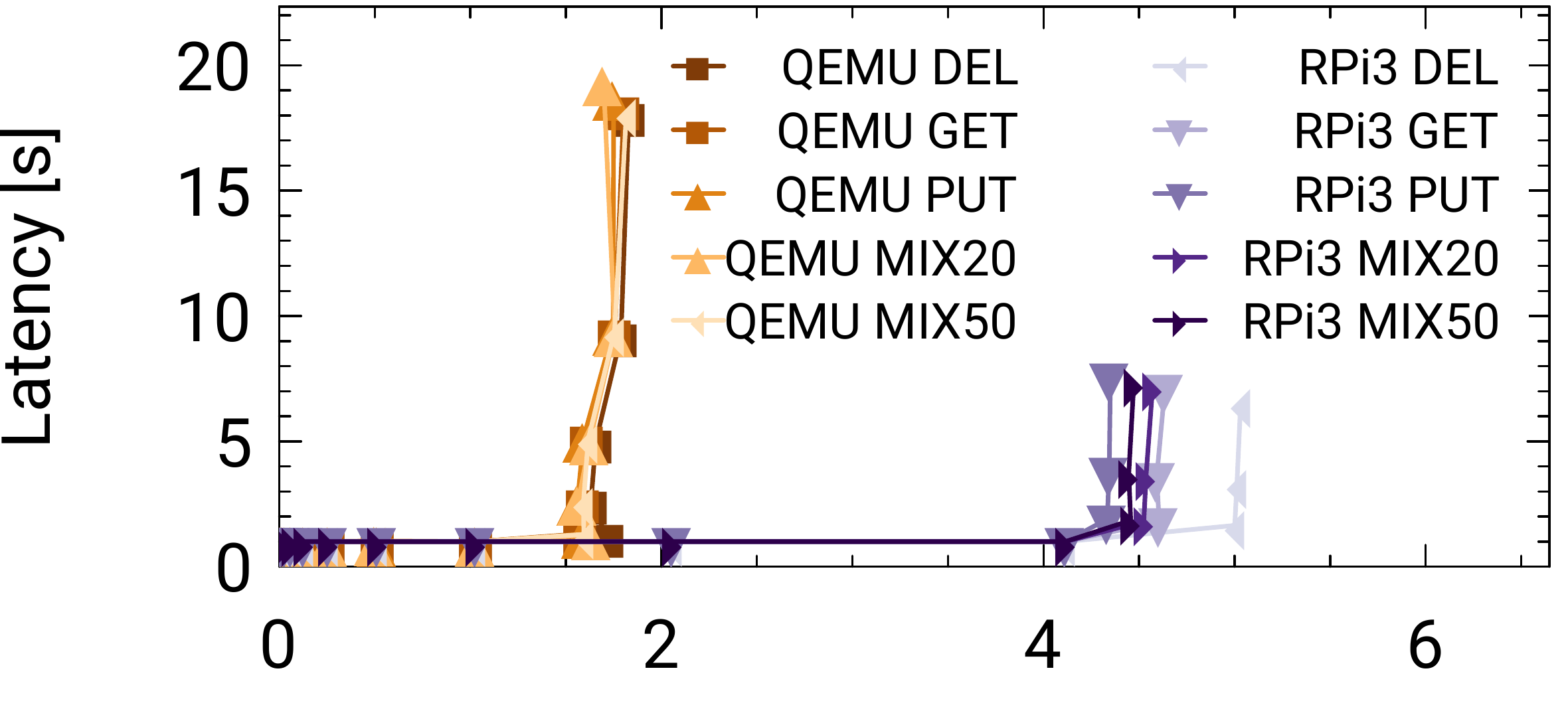}%
    \captionsetup{skip=0pt}
    \caption{Partially shared memory}%
    \label{fig:psm}
  \end{subfigure}%
  \quad%
  \begin{subfigure}[t]{0.46\textwidth}
    \includegraphics[trim={100px 30px 0 205px},clip,width=\linewidth]{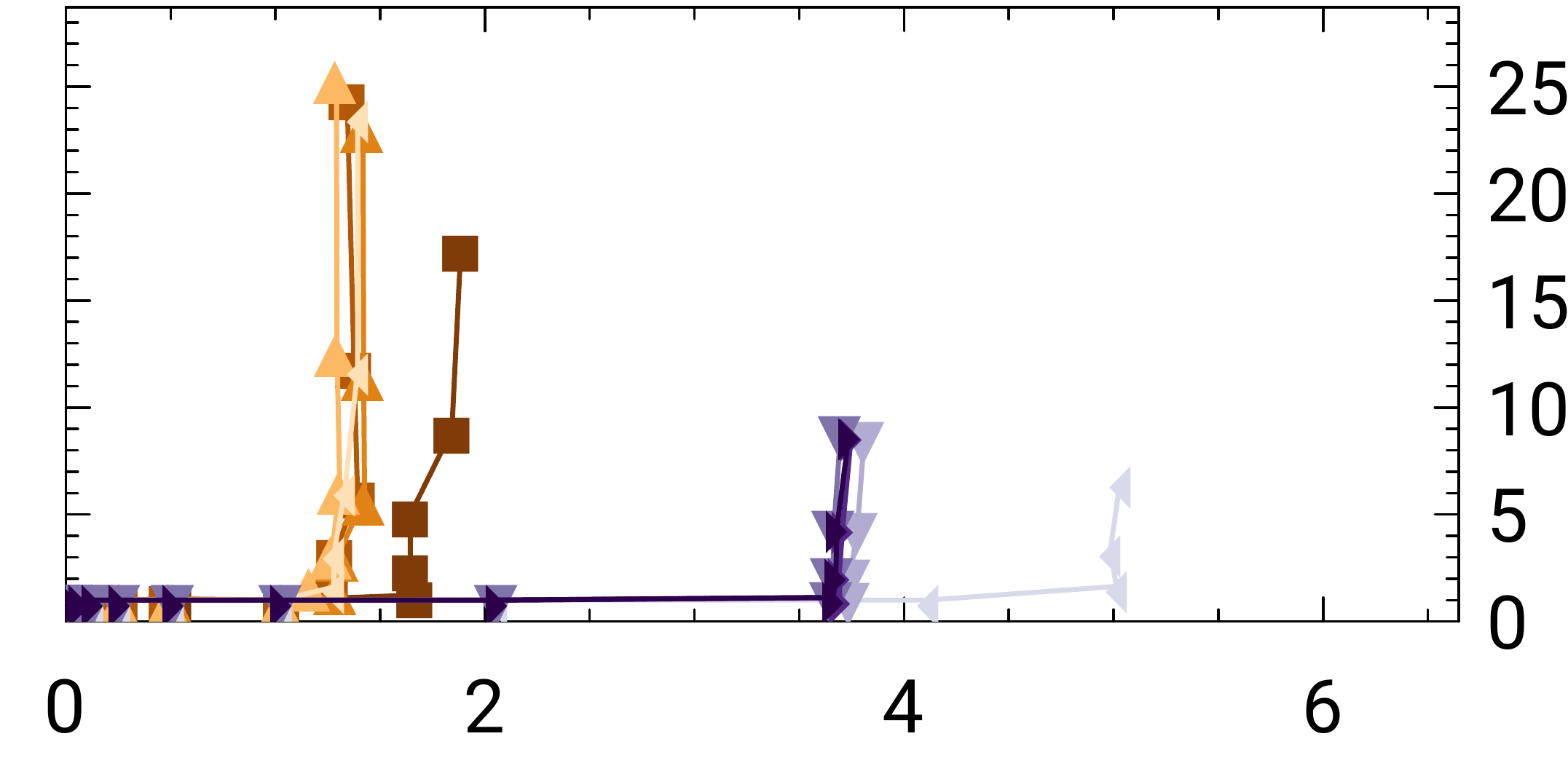}%
    \captionsetup{skip=0pt}
    \caption{Temporarily shared memory}%
    \label{fig:tsm}
  \end{subfigure}

  \begin{subfigure}[t]{0.50\textwidth}
    \includegraphics[trim={5px 0 35px 175px},clip,width=\linewidth]{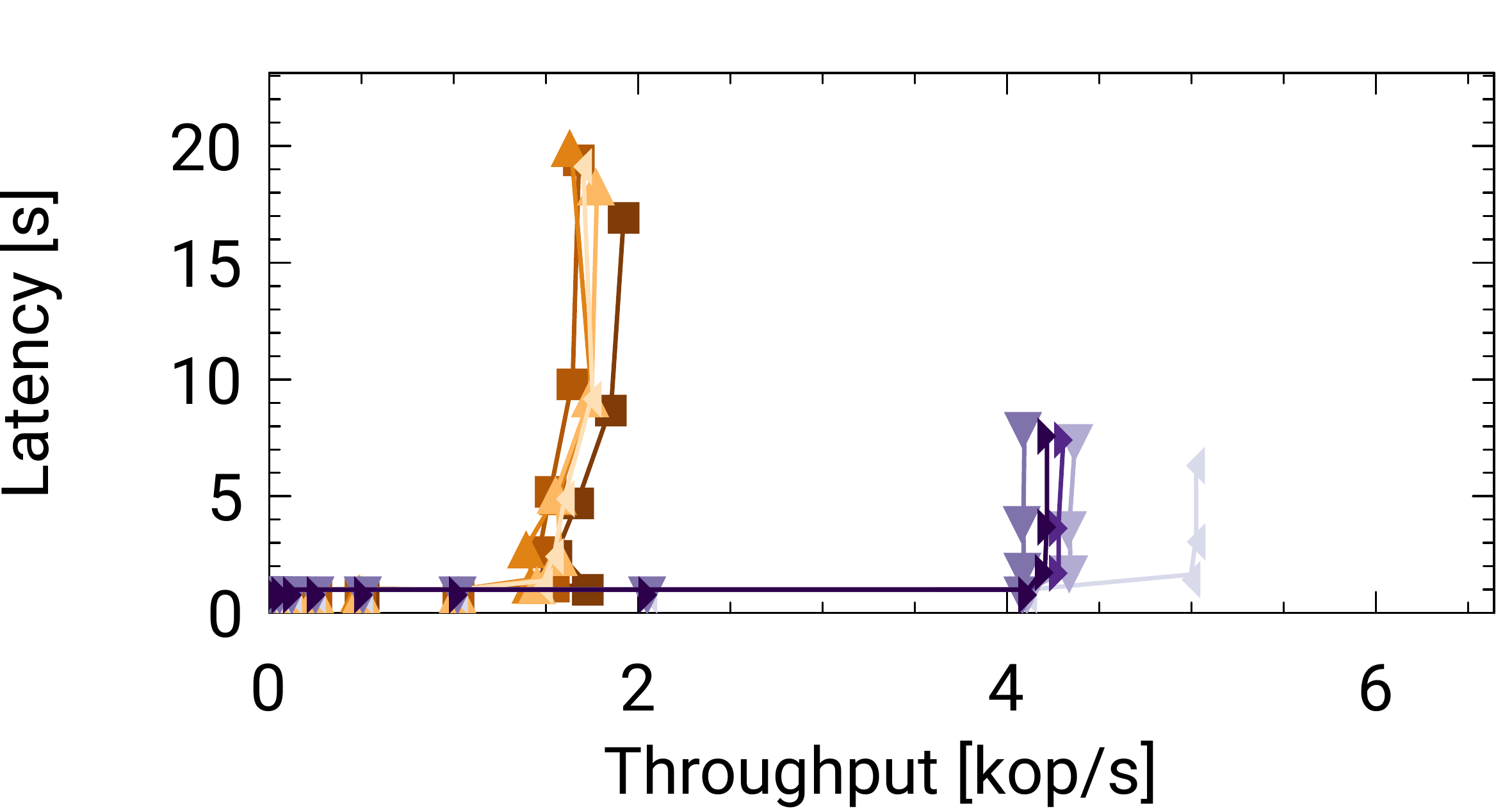}%
    \caption{Whole shared memory}%
    \label{fig:wsh}
  \end{subfigure}%
  \quad%
  \begin{subfigure}[t]{0.46\textwidth}
    \includegraphics[trim={100px 0 0 175px},clip,width=\linewidth]{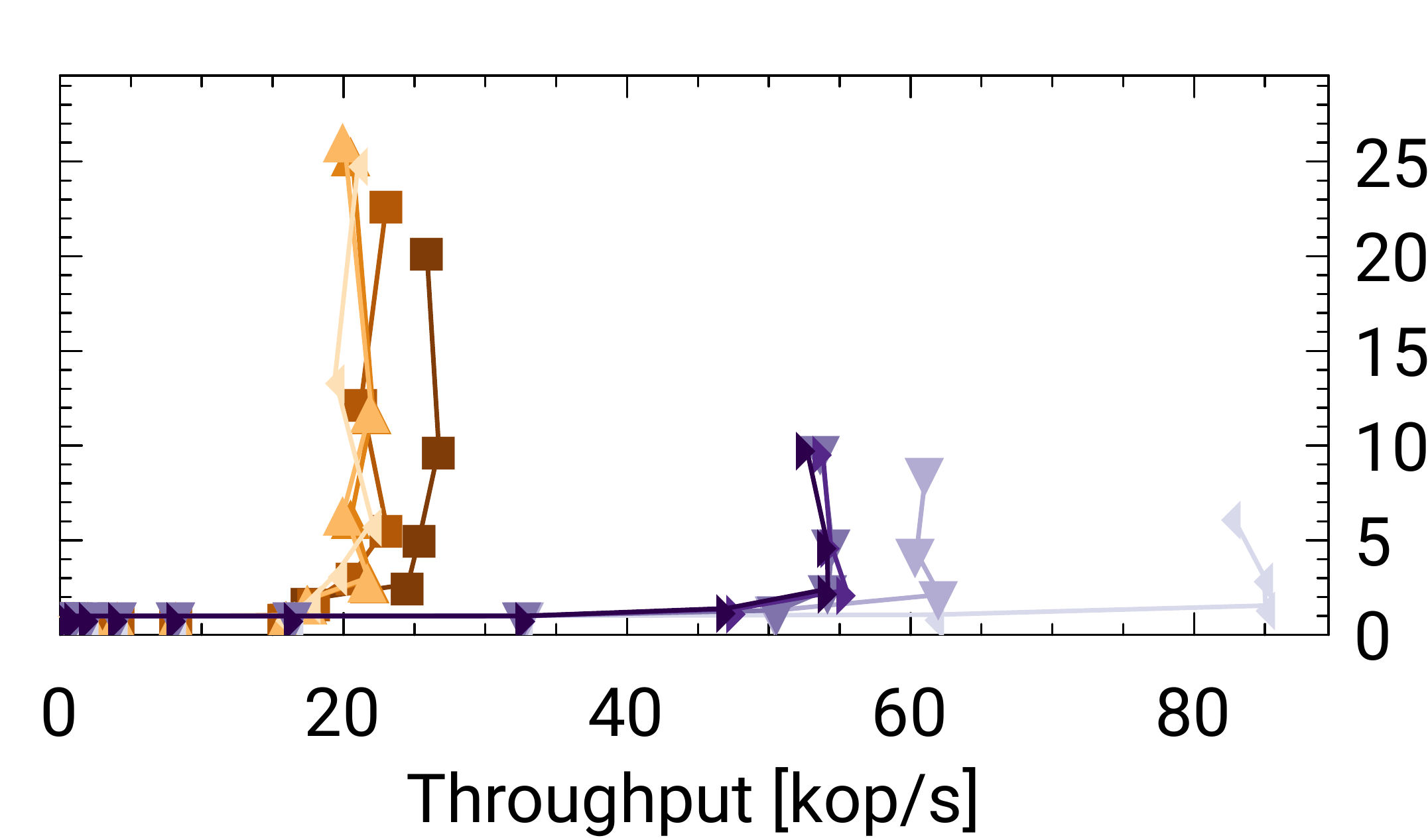}%
    \caption{CA in the REE}%
    \label{fig:ca}
  \end{subfigure}
  \caption{Throughput-latency plots for different kinds of shared memory.}%
  \label{fig:evalmemory}
\end{figure*}

\textbf{Power masurement.} To measure the power consumption of the two platforms, we connect the Dell PowerEdge server to a LINDY iPower Control 2x6M \emph{power distribution unit} (PDU)~\cite{lindy:pdu} and the Raspberry Pi 3B to an Alciom PowerSpy2~\cite{alciom:powerspy}.
The LINDY PDU provides a HTTP interface queried up to every second with a resolution of 1 W and a
precision of 1.5\%.
Alciom PowerSpy2 devices rely on Bluetooth channels to transfer the collected metrics. %
Both measuring devices collect voltage, current and power consumption in real time.

\textbf{Memory Bandwidth.}
We use an existing key-value store TA~\cite{dais19} to evaluate the overhead of the different types of shared memory.
The hash-table at the core of the key-value store uses separate chaining for collision resolution and implements modular hashing.
The \gp specification defines three different types of shared memory: \emph{whole} (an entire memory area), \emph{partial} (a subset of an entire memory area with a specified offset), and \emph{temporarily} (a memory area within the REE with an optional offset).
The temporarily shared memory area is only shared with the TA for the duration of the TEE method invocation; the two others get registered and unregistered with the TEE session.
The key-value store supports common operations such as \texttt{DEL}, \texttt{GET} and \texttt{PUT} on key-value pairs.
We benchmark each operation in isolation as well as combining \texttt{GET} and \texttt{PUT} operations (\texttt{MIX}ed benchmark).
The benchmarks operate as follows: for whole and partially shared memory, the CA will request a shared memory region of \SI{512}{\kibi\byte} from the TEE and fills it with random data from \texttt{/dev/urandom}.
With temporarily shared memory, the CA will allocate a \SI{512}{\kibi\byte} buffer and initialize it similarly with random data.
Before invoking a key-value operation a chunk size of \SI{1}{\kibi\byte} is selected as data object at a random offset in the shared memory respectively buffer.
The random offset is then used as key
and every operation is timed using \texttt{CLOCK\_MONOTONIC}.\footnote{Manual page: \texttt{man time.h}}
During the benchmark 256 operations are issued at a fixed rate between $1$ and $32768$ operations per second. %
\autoref{fig:evalmemory} shows the throughput-latency plots for each type of shared memory as well as for running the key-value store as a CA in the REE.

Compared to the Raspberry Pi, the results on QEMU are predominantly superposed and only achieve about half the throughput.
We believe this is due to an I/O bound from the \arm instruction and \tz emulation using QEMU.
We further observe with QEMU that the \texttt{DEL} benchmark for temporarily shared memory (\autoref{fig:tsm}) and as CA (\autoref{fig:ca}) is clearly distinguishable from the other benchmarks.
On the Raspberry Pi platform the graphs are well separated and ranked according to our expectations (lowest to highest throughput): \texttt{PUT}, \texttt{MIX50}, \texttt{MIX20}, \texttt{GET}, and \texttt{DEL}.
The \texttt{PUT} operation has the lowest throughput because of memory allocation, memory copy and object insertion in the TA.
The \texttt{GET} operation looks up the data object and copies it to the shared memory resulting in a higher throughput than the \texttt{PUT} operation.
The mixed benchmarks show a similar behavior: the higher the \texttt{PUT} ratio, the lower the throughput.
Hence, the \texttt{MIX50} ($50\%$ \texttt{PUT} operations) has a lower average throughput than \texttt{MIX20}.
The \texttt{DEL} operation avoids any time intensive memory operation and only has to free a data object after looking it up in the store.
An interesting observation is made when comparing the memory throughput of the benchmarks executed in the REE against the benchmarks executed in the TEE.
Key-value store operations executed inside TAs experience a 12$\times$-14$\times$ overhead with QEMU and a 12$\times$-17$\times$ overhead on the Raspberry Pi.
This overhead is due to the world and context switches associated to TA method invocations.

\begin{figure*}[t]
  \centering
  \begin{subfigure}[t]{0.48\textwidth}
    \includegraphics[trim={0px 0px 0px 205px},clip,width=\linewidth]{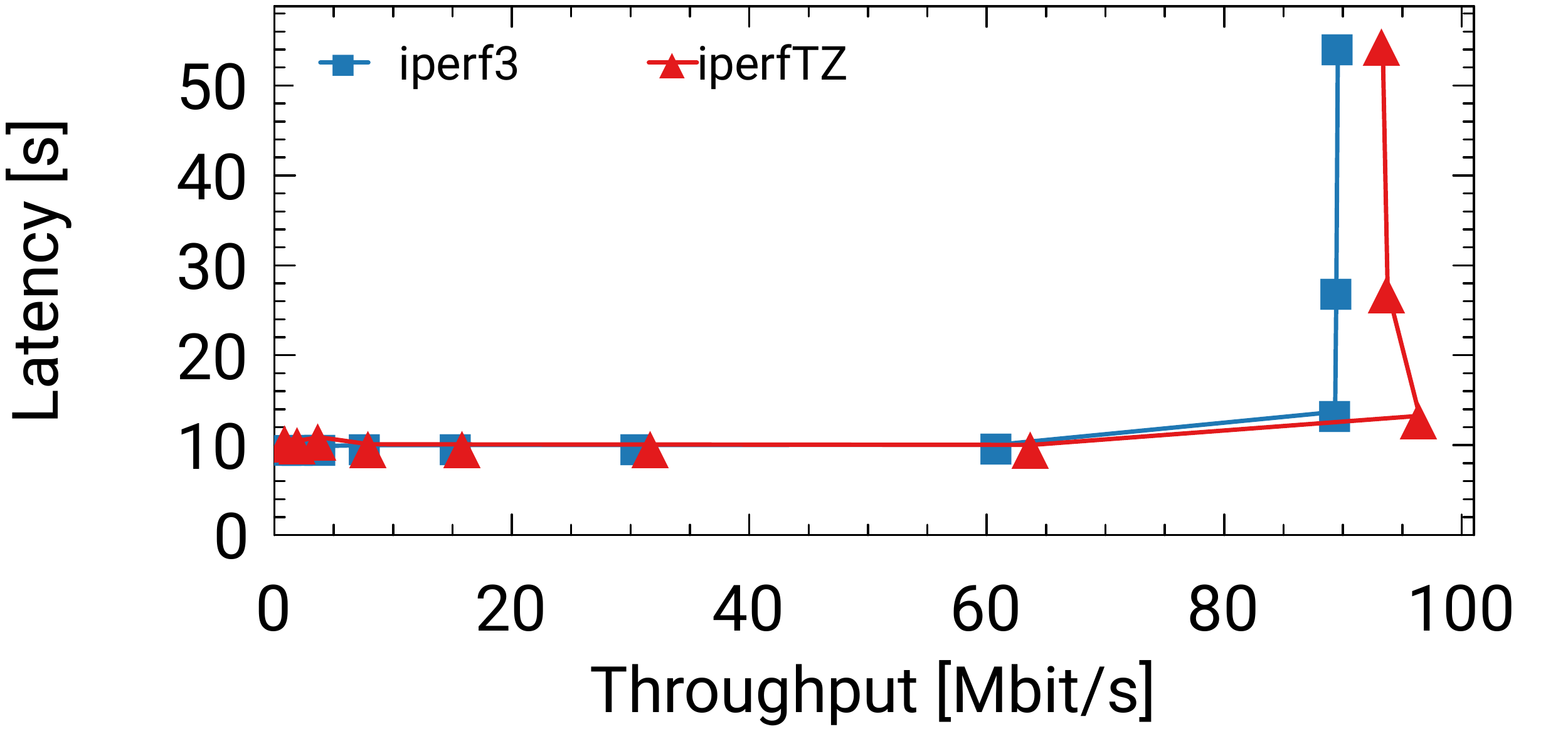}%
    \caption{Raspberry Pi}
  \end{subfigure}%
  \quad%
  \begin{subfigure}[t]{0.48\textwidth}
    \includegraphics[trim={0px 0 0 205px},clip,width=\linewidth]{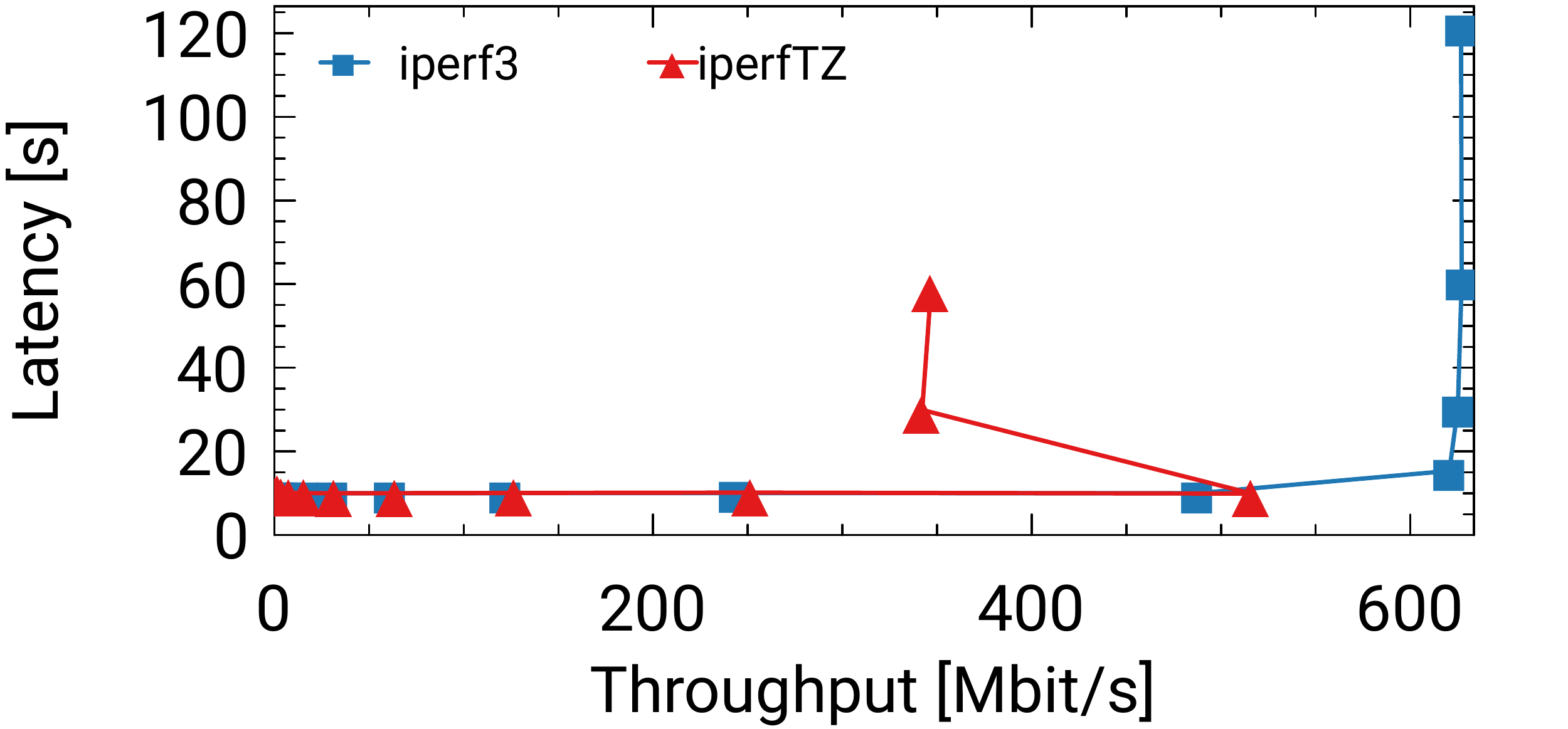}%
    \caption{QEMU}
  \end{subfigure}
  \caption{TCP network throughput measurements for \SI{128}{\kibi\byte} buffer sizes.}%
  \label{fig:evalthroughput}
\end{figure*}

\textbf{Network Bandwidth.}
This micro-benchmark compares the network throughput measured with \sys in \optee to the network throughput measured with \texttt{iperf3} in Linux.
We deploy both programs with the same set of parameters, \ie \SI{128}{\kibi\byte} socket and dummy buffer sizes.
Upon each iteration the bit rate is doubled starting at \SIrange[per-mode=symbol]{1}{512}{\mega\bit\per\second}. %
It should be noticed that TAs are by default limited to \SI{1}{\mebi\byte} of memory during runtime.
For this reason we do not allocate more than \SI{512}{\kibi\byte} for the dummy data on the TA's heap.
Linux has two kernel parameters which limit the maximum size of read and write socket buffers: \texttt{/proc/sys/net/core/rmem\_max} and \texttt{/proc/sys/net/core/wmem\_max}.
These kernel parameters controlling the socket buffer size limit can be changed at runtime using \texttt{sysctl}, in order to allocate larger socket buffers.

\sys is generally exceeding on both setups the network throughput of \texttt{iperf3}.
On the Raspberry Pi 3B we cannot observe any degradation of the network throughput due to an overhead from frequent world switches.
This result does not come as a surprise.
The memory bandwidth benchmark operates at a throughput of several hundred \si{\mega\byte\per\second}, while the network bandwidth benchmark operates at about \SI{10}{\mega\byte\per\second}.
There is a gap of one order of magnitude in throughput between the two benchmarks, which we assume to be sufficient for the overhead not to arise.
However, on QEMU we observe a serious degradation of the network throughput, when trying to achieve \si{\giga\bit\per\second} bit rate with \optee.
Remarkably, high throughput rates are strongly affected by the world switching overhead, even degrading beyond unaffected throughput rates.
Our measurements indicate that network throughput beyond \SI{500}{\mega\bit\per\second} is affected by a \SI{1.8}{\times} world switching overhead, almost halving the network throughput.

\begin{figure*}[t]
  \centering
  \begin{subfigure}[t]{0.48\textwidth}
    \includegraphics[trim={65px 15px 35px 305px},clip,width=\linewidth]{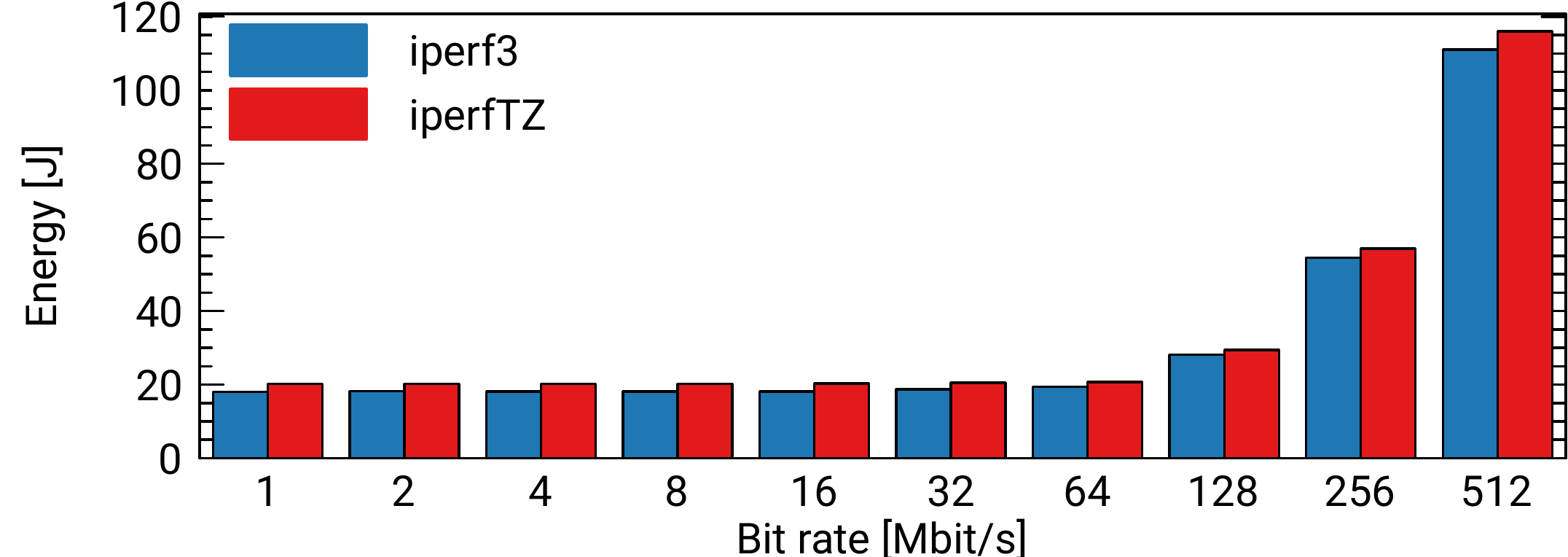}%
    \caption{Raspberry Pi}%
    \label{fig:rpie}
  \end{subfigure}%
  \quad%
  \begin{subfigure}[t]{0.48\textwidth}
    \includegraphics[trim={65px 15px 35px 305px},clip,width=\linewidth]{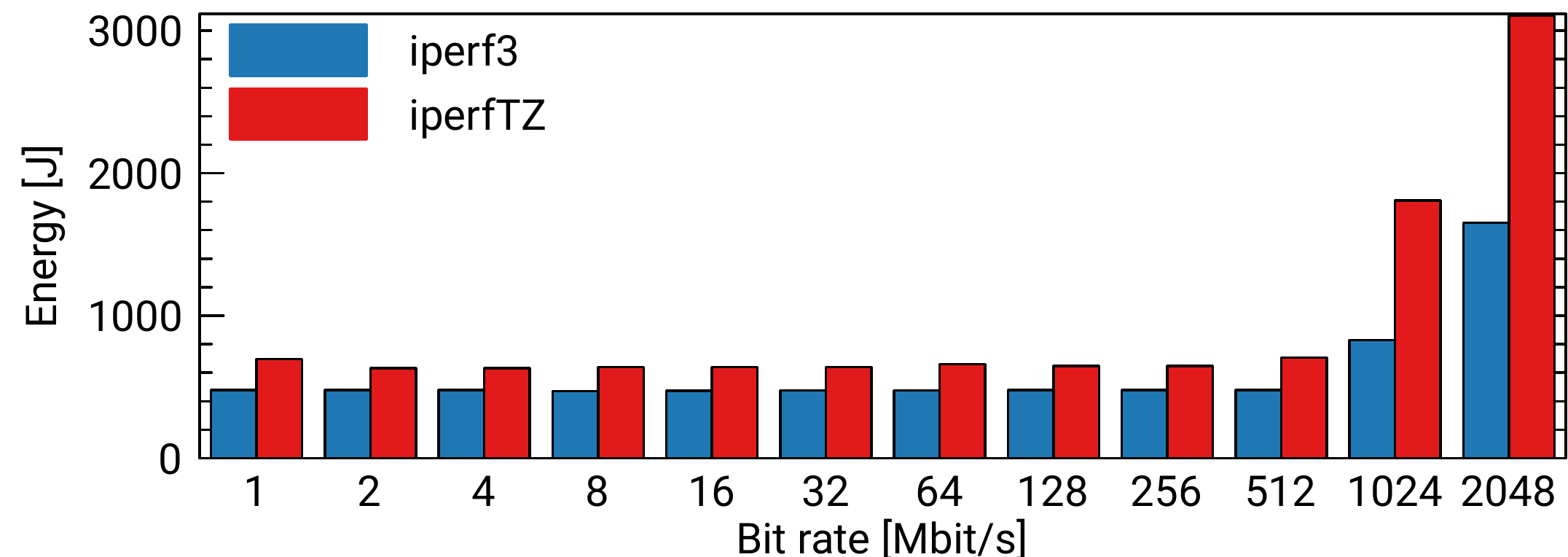}%
    \caption{QEMU}%
    \label{fig:qemue}
  \end{subfigure}
  \caption{Energy consumption during TCP network throughput measurements. Bit rates on the x-axis are given in logarithm to base 2.}%
  \label{fig:evalenergy}
\end{figure*}

\textbf{Energy.}
During the network bandwidth benchmark, we recorded the power consumed by both setups.
The LINDY iPower Control and the Alciom PowerSpy2 both record the timestamp as Unix time in seconds and the instantaneous power in watts.
We use those units to execute a numerical integration over time using the trapezoidal method to obtain the total energy consumed by both setups during a benchmark run.
\autoref{fig:evalenergy} shows these results.
The total energy on the y-axis (in joule) is consumed by the device while executing a benchmark run for a specific bit rate on the x-axis (as binary logarithmic scale in \si{\mega\bit\per\second}).
On the Raspberry Pi (\autoref{fig:rpie}) we observe that before reaching saturation, \sys is consuming about \SI{2}{\joule} (\SI{11}{\percent}) more than \texttt{iperf3}.
In the highly saturated range, the energy doubles with the throughput.
However, with QEMU (\autoref{fig:qemue}), the energy difference between the execution in the REE and the TEE is significant.
Given that QEMU is running on an energy-demanding and powerful server, \sys consumes about \SI{173}{J} (\SI{36}{\percent}) more before the overhead arises than \texttt{iperf3} in the REE. 
We can clearly attribute this additional energy consumption observed on both setups to the execution of \sys in the TEE.
Certainly, the world switching overhead also contributes to an increase of the energy consumption with QEMU.
By assuming a similar behavior for the energy consumption on QEMU as in the saturated range on the Raspberry Pi, we obtain a \SI{1.6}{\times} energy overhead due to world switching.
\section{Related Work}
\label{sec:rw}
There exists a plethora of network benchmarking and tuning tools.
We note that the implementation of \sys is heavily inspired by the well-known \texttt{iperf} tool.
In this sense, \sys support a subset of its command-line parameters, for instance to facilitate the execution of existing benchmarking suites.\footnote{Full compatibility with \texttt{iperf} would require substantial engineering efforts that we leave out of the scope of this work.}

The \texttt{ttcp} (Test TCP) tool was one of first programs implemented to measure the network performance over TCP and UDP protocols.
Lately, it has been superseded by \texttt{nuttcp}.\footnote{See~\autoref{fn:nuttcp}}
A tool with similar features is \texttt{netperf}.\footnote{See~\autoref{fn:netperf}}
Unlike the aforementioned tools,
\texttt{tcpdump}\footnote{\url{https://www.tcpdump.org}, accessed on 30.07.2019} is a packet analyzer that captures TCP packets being sent or received over a network.
\sys does not provide packet analysis tools. 
Instead, it does offer client and server-side measurements both for TCP and UDP datafows.
More recently, \texttt{iperf} integrated most of the functionalities of \texttt{ttcp}, extending it with multi-threading capabilities (since \texttt{iperf} v2.0) and allowing bandwidth measurements of parallel streams.
While it would be possible to provide similar support in \sys, the execution of code inside the TAs is currently single-threaded, hence limiting the achievable outbound throughput.
The most recent version of \texttt{iperf} (v3.0) ships a simplified (yet single-threaded) implementation specifically targeting non-parallel streams.
Flowgrind\footnote{\url{www.flowgrind.net}, accessed on 30.07.2019} is a distributed TCP traffic generator.
In contrast, \sys follows a client-server model, with traffic generated between a server and a TA.
StreamBox-TZ~\cite{park2019streambox} is a stream analytics engine, which processes large IoT streams on the edge of the cloud.
The engine is shielded from untrusted software using \tz.
Similar to \sys, StreamBox-TZ runs on top of \optee in a TA.
Yet, \sys does not process data streams but can generate and measure network performance of those streams.

To summarize and to the best of our knowledge, \sys is the first tool specifically designed to run as a TA for \tz that can measure the achievable network throughput for such applications.

\section{Conclusion and Future Work}
\label{sec:conclusion}
The deployment of TAs is becoming increasingly pervasive for the management and processing of data over the network.
However, due to constraints imposed by the underlying hardware and runtime system, network performance of TAs can be affected negatively.
\sys is a tool to measure and evaluate network performance of TAs for \arm \tz, a widely available TEE on embedded, IoT and mobile platforms.
We implemented the \sys prototype on top of \optee and we evaluated it on the Raspberry Pi platform.
Our experimental results highlight performance and energy trade-offs deployers and programmers are confronted with both on hardware and emulated environments.
We believe the insights given by our work can be exploited to improve design and configuration of TEEs for edge devices handling real-world workloads for TAs.  

We intend to extend our work to support different types of sockets (\eg, datagram sockets) and to leverage on-chip cryptographic accelerators.
This would allow us to provide TLS-like channels for TAs, a feature that has not yet been implemented in \optee.
Finally, we aim for supporting various kinds of TEEs, especially in the context of embedded platforms and SoC, such as Keystone\footnote{\url{https://keystone-enclave.org}, accessed on 30.07.2019} for RISC-V processors.

\section*{Acknowledgments}
The research leading to these results has received funding from the European Union's Horizon 2020 research and innovation programme under the LEGaTO Project (\href{https://legato-project.eu/}{legato-project.eu}), grant agreement No~780681.

{
\footnotesize
\bibliographystyle{splncs04}
\bibliography{biblio}
}

\end{document}